\documentclass[12pt]{iopart}
\usepackage{graphicx}

\begin{document}
\title[Planar hybrid Ba-122 SNS' thin film junctions]{Planar hybrid superconductor-normal metal-superconductor thin film junctions based on BaFe$_{1.8}$Co$_{0.2}$As$_2$}
\author{S D\"oring$^1$, S Schmidt$^1$, F Schmidl$^1$, V Tympel$^1$, S Haindl$^2$, F Kurth$^2$, K Iida$^2$, I M\"onch$^2$, B Holzapfel$^2$ and P Seidel$^1$}
\address{$^1$ Institut f\"ur Festk\"'orperphysik, Friedrich-Schiller-Universit\"at Jena, Helmholtzweg 5, 07743 Jena, Deutschland}
\address{$^2$ Institut f\"ur metallische Werkstoffe, IFW Dresden, Postfach 270116, 01171 Dresden, Deutschland}
\ead{sebastian.doering.1@uni-jena.de}

\begin{abstract}
To investigate the transport properties of iron based superconductors, we prepared planar hybrid superconductor-normal metal-superconductor (SNS') thin film junctions with BaFe$_{1.8}$Co$_{0.2}$As$_2$ as base electrode. As counter electrode we used a lead indium alloy, while the normal metal layer was formed by thin gold films. Temperature dependent measurements of the electrical conductivity were strongly influenced by the properties of the electrodes. We developed a junction structure that allows us to characterize the electrodes, too, including the behavior of their normal state resistance in order to correct their influences on the conduction spectra. The corrected conductivity of the junction was described within an extended BTK-model and shows a behavior dominated by Andreev reflexion.
\end{abstract}

\maketitle

\section{Introduction}
To investigate the superconducting properties of iron based superconductors Andreev reflection studies are a potent tool. By examining SNS' thin film junctions, fundamental properties like the energy gap and order parameter symmetries can be derived. Recent experiments on the BaFe$_2$As$_2$ mother compound (Ba-122) employ different kinds of techniques for acquiring information about the order parameter and are reviewed in \cite{Seidel2011}. The most commonly used compouds of Ba-122 are the electron doped BaFe$_{2-x}$Co$_{x}$As$_2$ and hole doped Ba$_{1-x}$K$_x$Fe$_2$As$_2$. In both, the examined number of gabs differs. For the K-doped Ba-122 there are results for two gaps \cite{Szabo2009,Samuely2009} as well es for a single gap \cite{Lu2010,Zhang2010}.

Also for the here investigated Co-doped Ba-122 there are different results. Measurements of the optical conductivity suggest a single gap with a $2\Delta$/$k_B T_c$ ratio of 2.1 \cite{Gorshunov2010}, while calorimetric investigations showed two gaps with values of 1.9 and 4.4, respectively \cite{Hardy2010}. Point contact measurements of the electrical conductance were performed by Samuely \etal \cite{Samuely2009} and Massee \etal \cite{Massee2009,Massee2010} via point-contact spectroscopy. Both reported a single gap with $2\Delta$/$k_B T_c$ equals 5.8 and 7.4, respectively. In contrast, other groups found two gaps with $2\Delta$/$k_B T_c$ ratios of 1.7 and 10.2 \cite{Park2011} and 3.3 and 6.6, respectively \cite{Teague2011}.

The main disadvantage of point-contact methods is the undefined and thus hardly reproducible junction barrier. By preparing planar junctions, we are able to vary the material used for the normal layer, their thickness and the junction area within one sample of BaFe$_{1.8}$Co$_{0.2}$As$_2$. Here we present preparation and measurements on such junctions.

\section{Preparation of the junctions}
\label{prep}
\begin{figure}[htbp]
\centering
\includegraphics[width=.49\columnwidth]{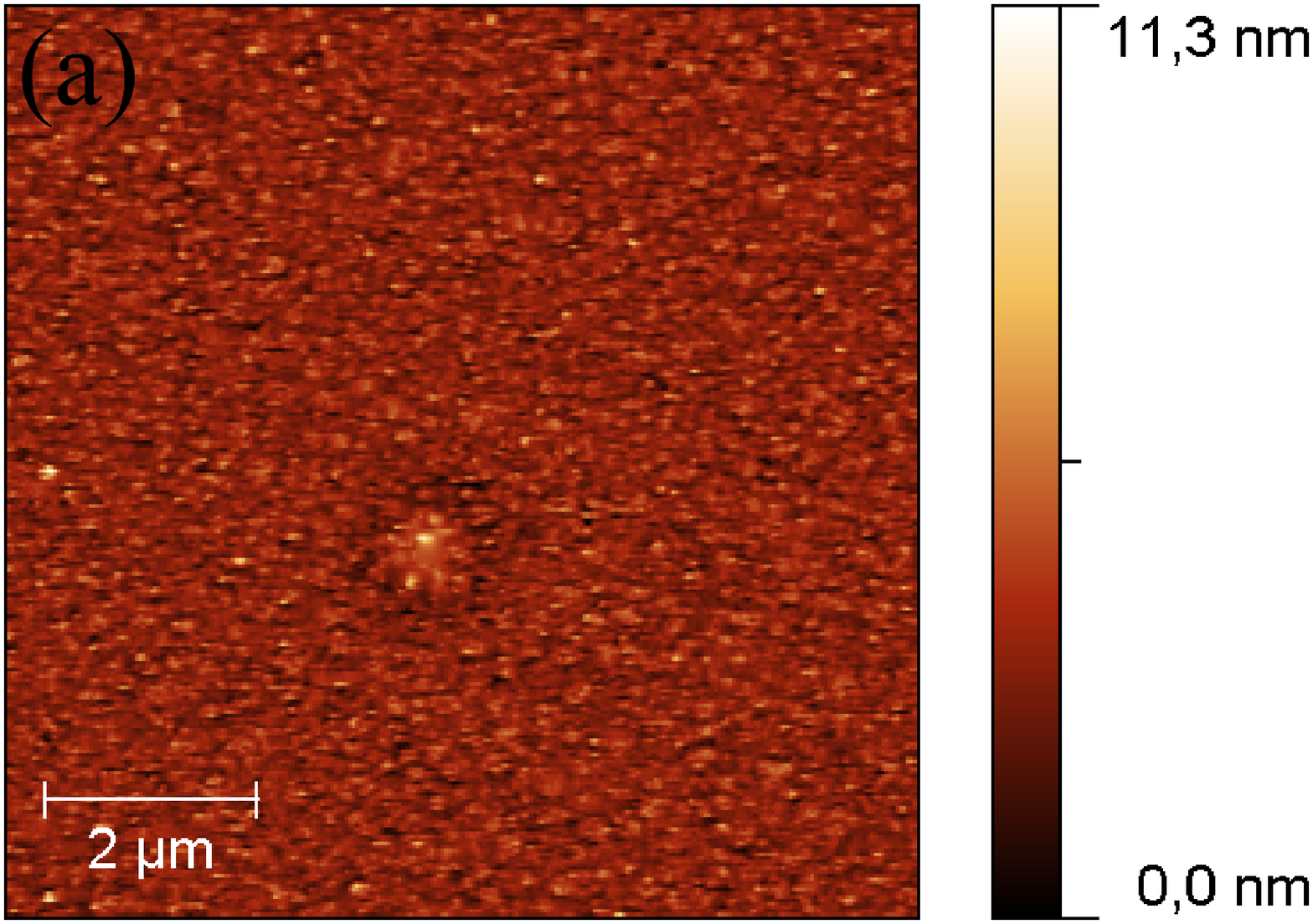}
\includegraphics[width=.49\columnwidth]{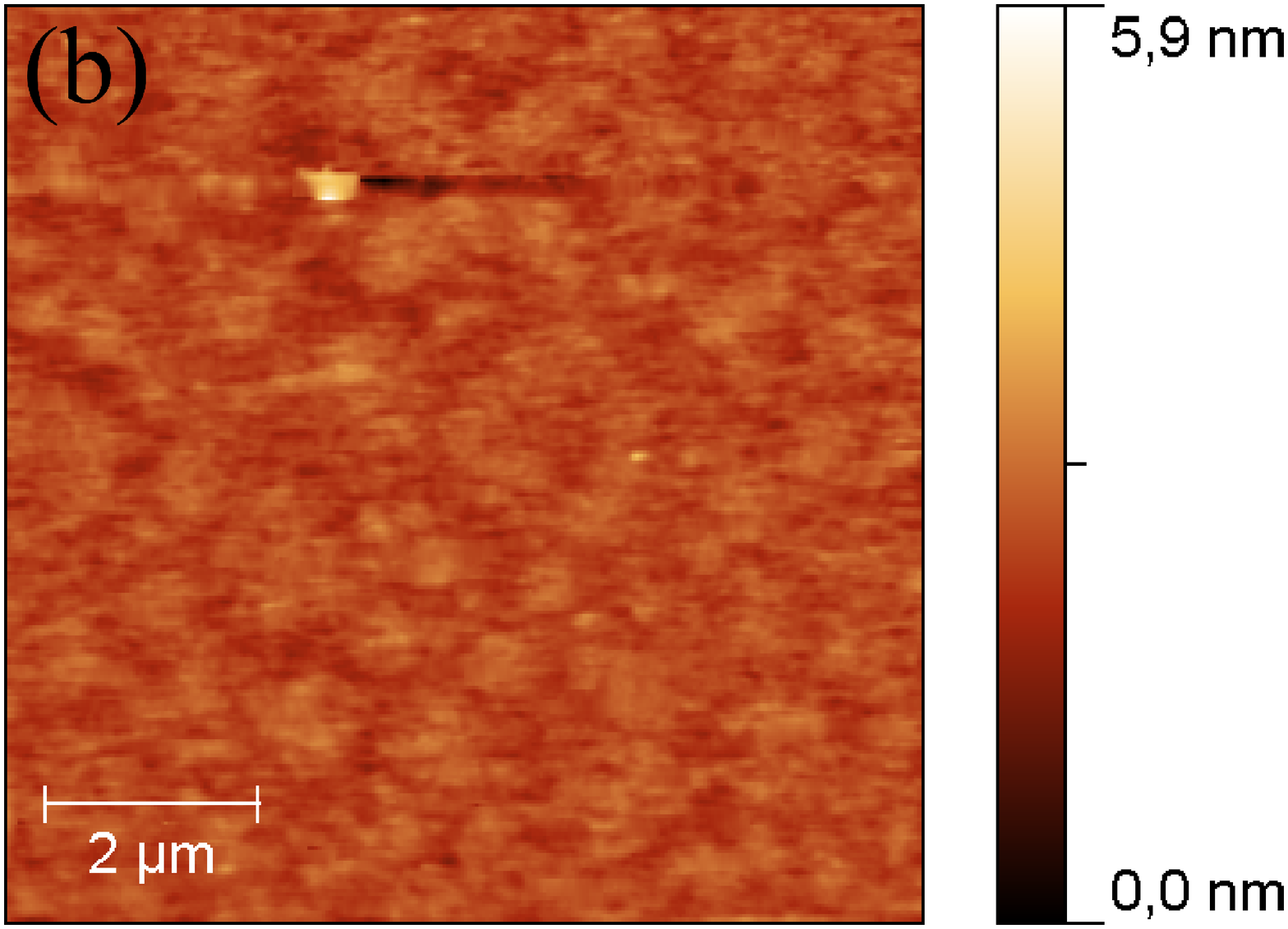}
\caption{\label{fig:AFM}Atomic force microscope images of the film surface of the Ba-122 film as deposited (a) and the Ba-122 covered with a $10$\,nm thick gold layer (b).} 
\end{figure}

To prepare the junctions, thin films of Ba-122 were used, which were fabricated by pulsed laser deposition (PLD) on a (La,Sr)(Al,Ta)O$_3$ (LSAT) substrate. The films had thicknesses of approximately $80$\,nm. Details of the deposition process can be found in \cite{Iida2010}. To cover the whole sample a gold layer with a thickness of $10$\,nm was sputtered. The surface showed a good quality with a root mean square (RMS) roughness of less than $1$\,nm before and after the sputter process (see \fref{fig:AFM}). The gold layer forms the barrier between the two superconducting electrodes as well as it avoids possible degradation of the Ba-122 by air, photo resists and other chemicals used in subsequent preparation steps. The thickness of this layer is tunable. For comparison, similar samples prepared with a $5$\,nm gold layer, show clear Josephson effects \cite{Schmidt2010}. By increasing the thickness of the gold layer up to $10$\,nm coupling between the superconducting electrodes and thus the critical Josephson current was suppressed. Therefore, samples with thicker barrier layers, more suitable for Andreev reflection studies were chosen in this study.

The Ba-122 base electrode was patterned using photolithography and ion beam etching (IBE) with argon. Optionally, the removed material could be filled by sputtering SiO$_2$, a procedure recommended for thicker films of Ba-122 to avoid cracks in the counter electrode near the edges of the base electrode. 

\begin{figure}[htbp]
\centering
\includegraphics[width=.49\columnwidth]{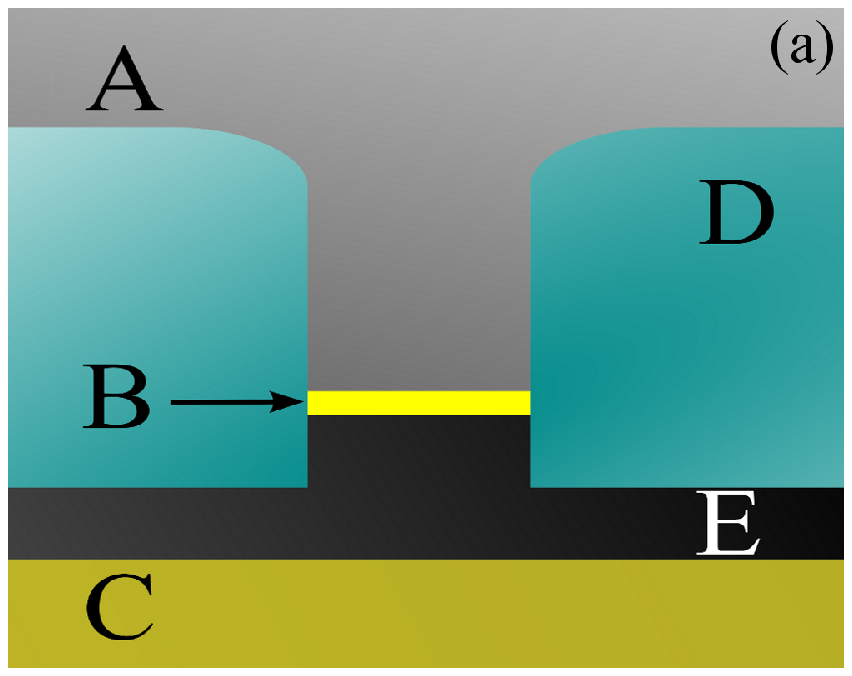}
\includegraphics[width=.49\columnwidth]{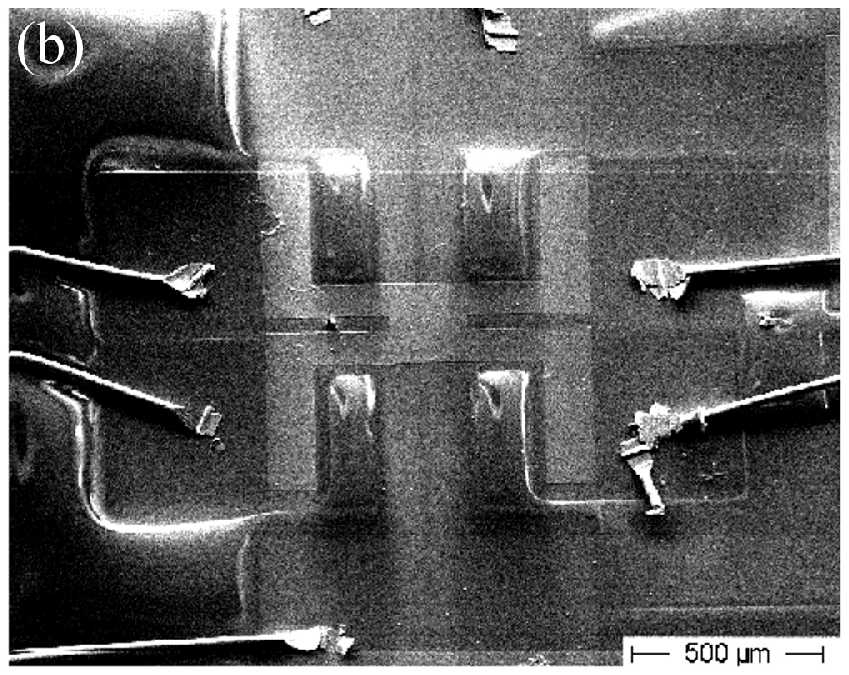}
\caption{\label{fig:Tunnel}(a) Schematic cross section of the junction based on photolithographic mask design. (b) Overview of a prepared junction acquired by scanning electron microscopy. A: PbIn counter electrode. B: Au barrier. C: LSAT substrate. D: SiO$_2$ framework. E: Ba-122 base electrode. The red arrow marks the junction area.}
\end{figure}

The preparation of the junction areas was performed following described steps. Firstly, we removed the gold and a small area of the Ba-122 layer around the later junction area via IBE. Secondly frameworks of SiO$_2$ with a thickness of $100\,nm$ were sputtered in order to insulate the electrodes against each other and to define the size of the junction. With our layout we are able to prepare ten junctions with areas between $100\,\mu$m $\times 100\,\mu$m and $3\,\mu$m $\times 3\,\mu$m on a single sample ($5$\,mm $\times 10$\,mm). Finally, a lead indium (PbIn) alloy was deposited by thermal evaporation for the counter electrode (thickness $\approx150\,nm$). The final structure is shown in \fref{fig:Tunnel}.

Our junction design precisely allows the determination of $T_c$ and $I_c$ for both electrodes separately as well as for the junction itself in four-point geometry. Therefore, it is possible to investigate the temperature dependent electrical properties for each electrode independently and determine their influence on the junction. All eight pads of one junction were contacted via ultrasonic bonding technique with gold wires (diameter $25\,\mu$m) providing the connection to the measurement equipment.

\section{Results and discussion}
\label{results}
Characterizing the junction and the electrodes requires information about the differential conductance in dependence of the applied voltage and the differential resistance of the biased current, respectively. For that, a $V$-$I$-characteristic was taken for each electrode as well as for the junction and the respective derivation was acquired numerically. The PbIn counter electrode has a critical temperature $T_c$ of $7.2$\,K. The differential resistance at $T=4.2$\,K is zero up to a critical current $I_c$ of $20$\,mA and a critical current density $j_c$ of $1.7\times10^{5}$\,Acm$^{-2}$, respectively. For $T\geq T_c$ it is independent from the biased current and slightly increases with temperature.
\begin{figure}[htbp]
\centering
\includegraphics[width=.8\columnwidth]{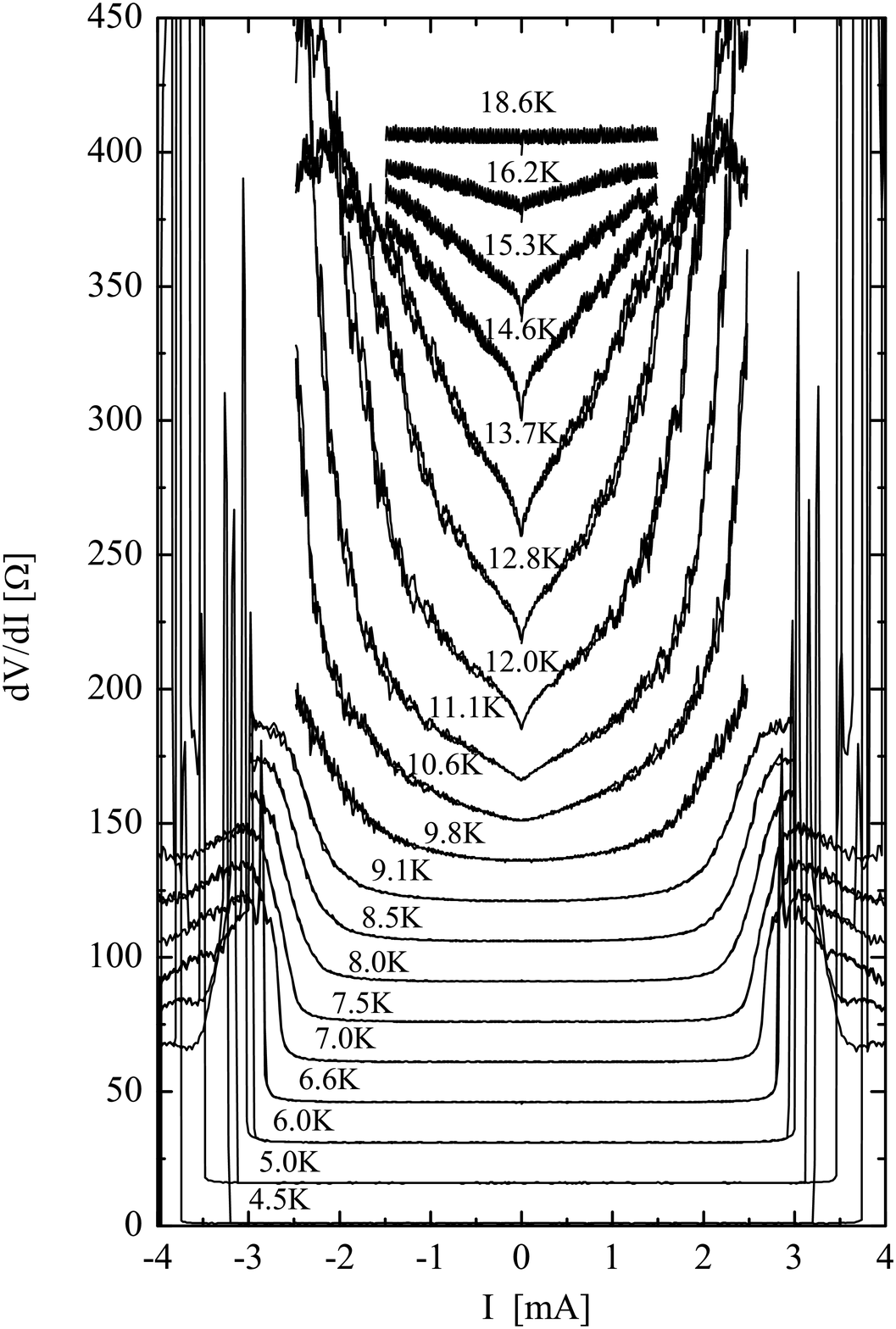}
\caption{\label{fig:Feas_Rdiff}Differential resistance versus current of the Ba-122 electrode at different temperatures. Each curve is shifted by $15\,\Omega$ against the one  below.}
\end{figure}

The pure Ba-122 base electrode behaves differently as it is shown in \fref{fig:Feas_Rdiff}. The $V$-$I$-curve shows hysteretic behavior up to $6.6$\,K, which is in contrast to measurements performed on microbridges \cite{Rall2011}. At $T=4.5$\,K the critical current is $(3.75\pm0.11)$\,mA, which corresponds to $j_c=(6.7\pm1.1)\times10^4$\,Acm$^{-2}$. At $T=11.1$\,K the critical current vanishes, which agrees with the $T_{c,0}$ of the $R$-$T$-curve measured with a bias current of about $10\,\mu$A. This is in good agreement with the values of the pure Ba-122 thin film \cite{Iida2010}. Thus, the preparation process described in \sref{prep} does not influence the properties of the Ba-122. By increasing the temperature the shape of the differential resistance changes. At temperatures above $10.6$\,K v-shaped behavior occurs for low currents instead of the u-shaped one at lower temperatures. This nonlinear current dependency will be discussed later in this article. At $T=18.6$\,K the Ba-122 thin film is in the normal state and the resistance reaches a constant value of $R_N=150\,\Omega$.

\begin{figure}[htbp]
\centering
\includegraphics[width=.49\columnwidth]{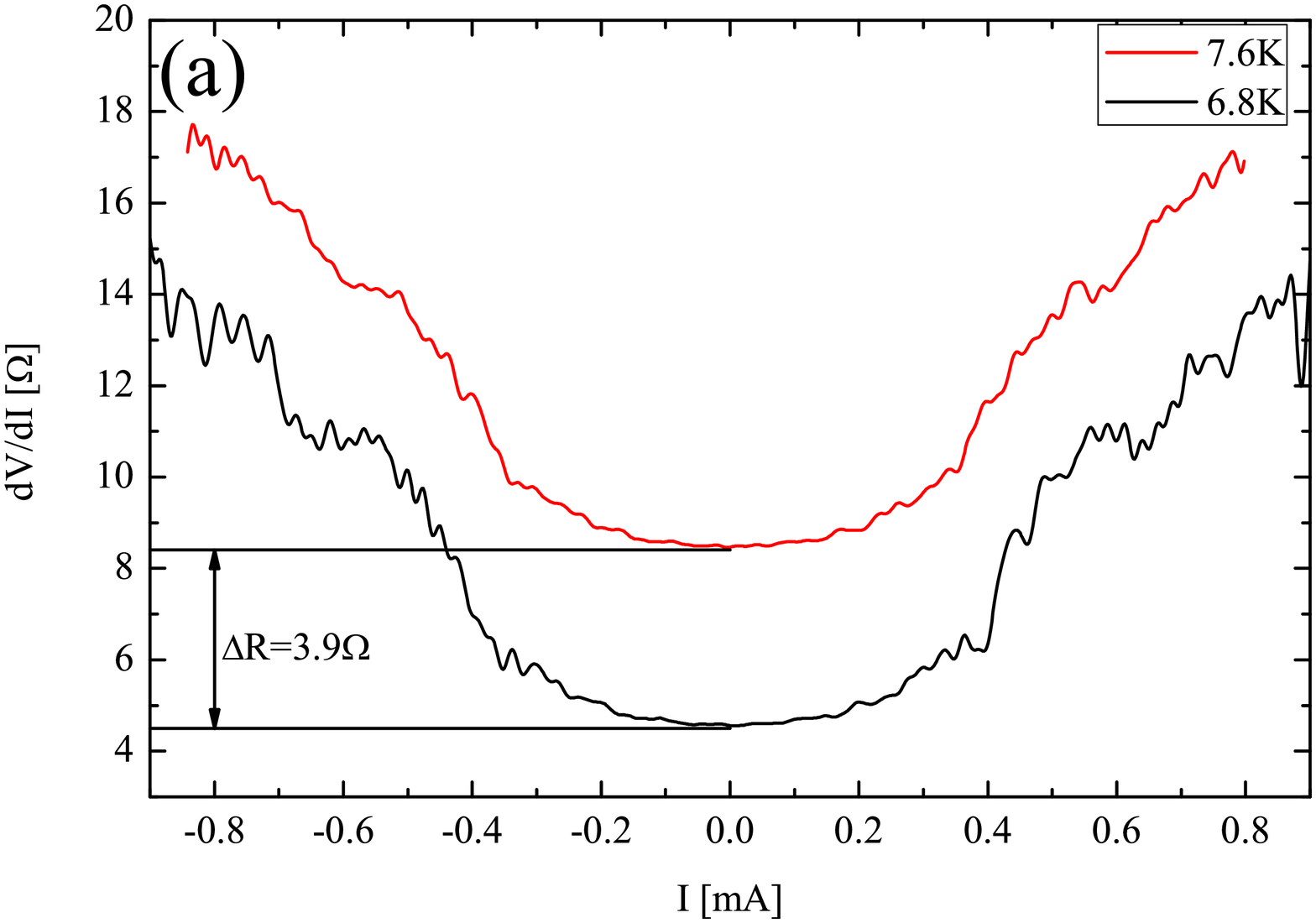}
\includegraphics[width=.49\columnwidth]{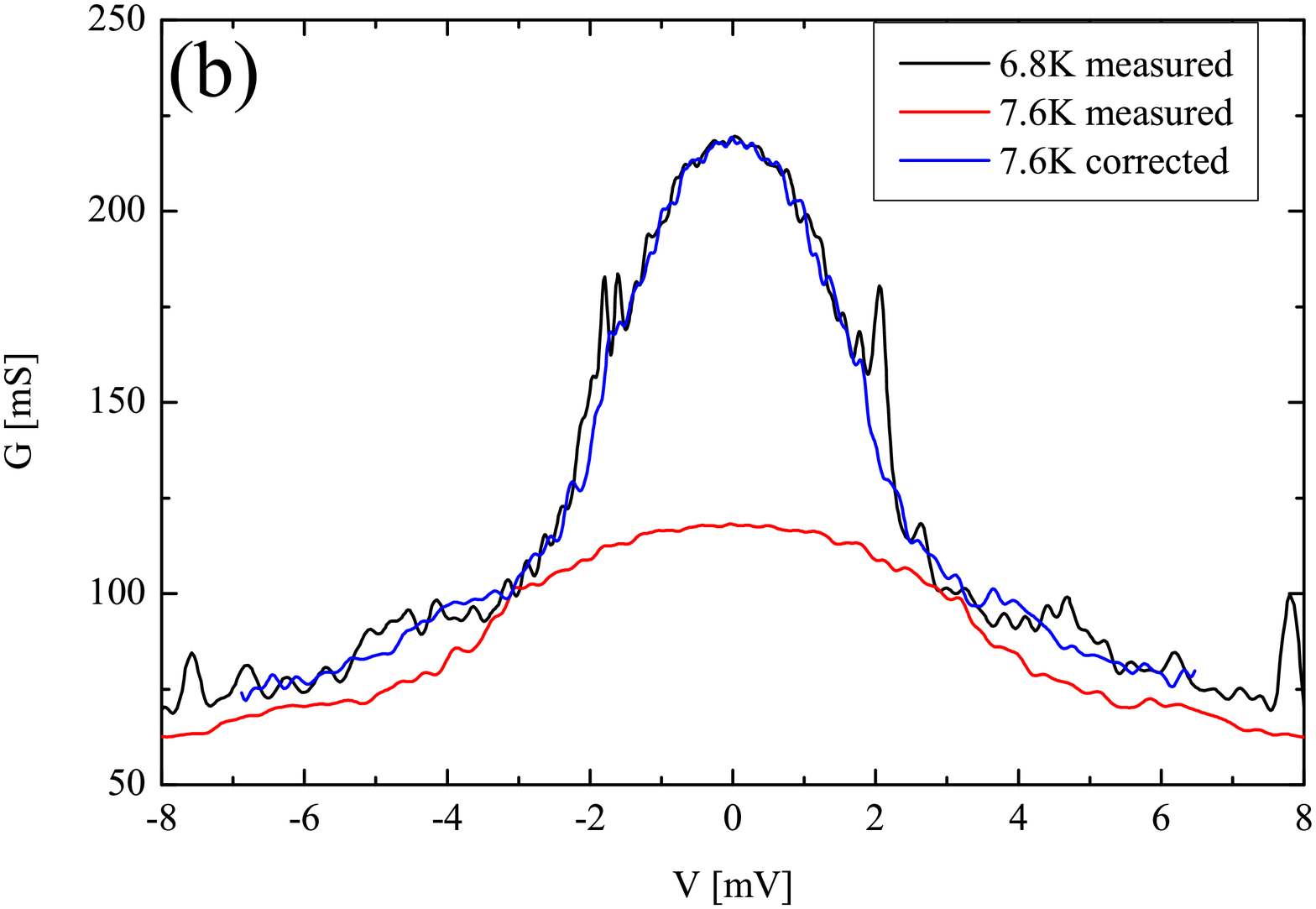}
\caption{\label{fig:Tu_Rdiff-G_T} (a) Differential resistance versus current of the junction at $6.8$\,K and $7.6$\,K. (b) Differential conductance versus voltage of the junction as measured at $6.8$\,K and as measured and corrected for $7.6$\,K.}
\end{figure}

In comparison, the differential resistance of a $3\,\mu$m $\times 3\,\mu$m junction is shown for temperatures of $6.8$\,K and $7.6$\,K, respectively (\fref{fig:Tu_Rdiff-G_T}(a)). The shape of both curves is very similar, but the differential resistance at $7.6$\,K is higher by an offset of about $3.9\,\Omega$. This is due to the normal state resistance of the PbIn counter electrode, which occurs for $T\geq7.2$\,K. If the counter electrode switches from superconducting to normal state, the measurement on the junction also switches from a four-point geometry to a three-point one. In this case the resistance of the electrode is in series to the resistance of the junction, and the total measured voltage can be written as sum of the junction voltage and the voltage at the counter electrode. The correct value of the voltage at the junction is then given to be:
\[V_{junction}=V_{measured}-R_{PbIn}\cdot I_{bias}.\]
It has also to be taken into account that the electrode resistance slightly increases with temperature. The result of this correction is shown in \fref{fig:Tu_Rdiff-G_T}(b). 
\begin{figure}[htbp]
\centering
\includegraphics[width=.8\columnwidth]{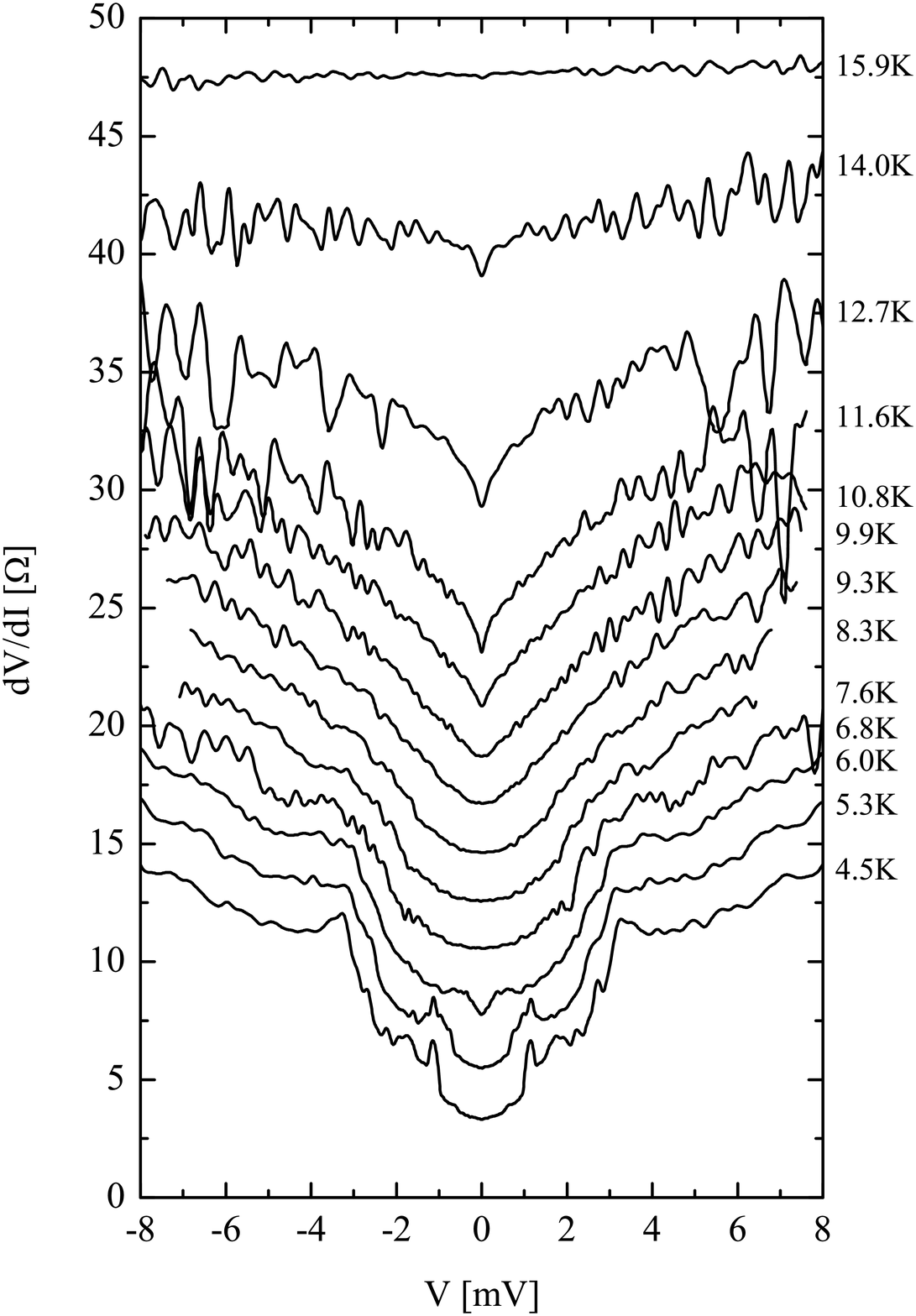}
\caption{\label{fig:Tunnel_Rdiff}Differential resistance versus voltage of the junction. The curves for $T\geq7.6$\,K are corrected as described in the text, while the other curves are shown as measured.}
\end{figure}

For the investigation of junctions, the differential conductance versus voltage is used. Subtraction of the electrode resistance clearly changes the shape of the conductance spectra, so that the curves for $T=6.8$\,K and $T=7.6$\,K look nearly equal. For reducing possible errors in the correction temperatures close to $T_c$ should be examined. This could be challenging due to high noise caused by temperature drift and fluctuation while measuring near $T_c$.

The Ba-122 base electrode also influences the measured spectra. As shown in \fref{fig:Tunnel_Rdiff} the nonlinear v-shaped behavior of the electrode, which was mentioned above, can be also seen in the spectra for $T\geq10$\,K. Due to the nonlinear dependence of the electrode resistance one can not, unlike for the counter electrode, derive the junction spectra from the measured one. To describe the conduction spectrum of a single superconductor-normal metal junction one has to use a BTK-model \cite{Blonder1982} with quasiparticle lifetime extension parameter $\Gamma$ \cite{Dynes1984,Plecenik1994}. A detailed description about the used model and possible extensions for iron-based superconductors (e.g. two-gap superconductivity) is given by Daghero and Gonelli \cite{Daghero2010}.

\begin{figure}[htbp]
\centering
\includegraphics[width=.8\columnwidth]{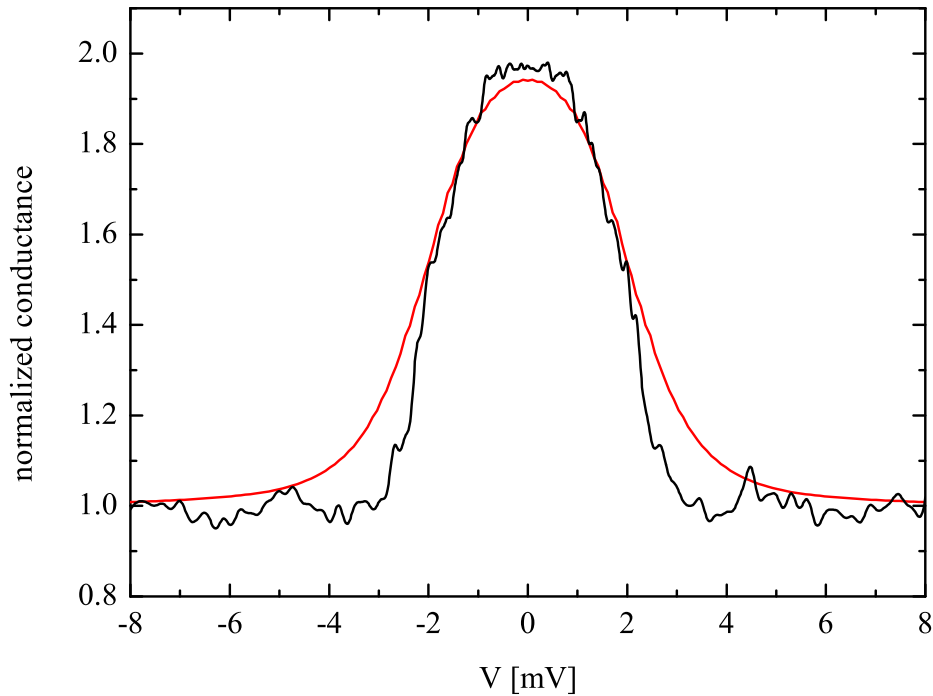}
	\caption{\label{fig:BTK_Fit} Normalized conductance of a junction at $T=6.7$\,K. The black curve shows the measured data, while the red line shows a fit within an extended BTK-model.}
\end{figure}

For the junctions presented here a description by the BTK-model is possible for temperatures lower than $10$\,K as it is shown in \fref{fig:BTK_Fit}, but despite a general agreement between the model and the measured data larger deviations are found in the range of medium voltages ($2$--$5$\,meV). The used parameters for this curve were $\Delta=(1.66\pm0.03)$\,meV, $Z=(0.07\pm0.04)$ and $\Gamma=(0.4\pm0.2)$\,meV. It can be seen, that the spectrum shows a feature of only a single gap. Also a performed comparison within a two gap model does not lead to a better fit with the measured data. For $T\geq10\,K$, the main features of the spectrum are caused by the Ba-122 electrode and not by the junction itself, which prevents a detailed analysis of the energy gap properties near its critical temperature.

\section{Summary}
To conclude, we developed a layout for SNS' junctions, which allows us to examine the temperature dependence of both electrodes and a SNS' junction itself independently from each other. We were able to measure and subsequently correct the influence of the normal state resistance of the PbIn counter electrode on the junction spectrum. Also we showed, that the behavior of the Ba-122 base noticeably influences the junctions spectra for $T\geq10\,K$. Finally, we described the conduction spectra of a junction within an extended BTK-model.

\ack
The authors would like to thank Veit Grosse for his contribution. This work was supported by the DFG within SPP 1458.

\end{document}